\documentclass[sigconf]{acmart}

\copyrightyear{2024}
\acmYear{2024}
\setcopyright{acmlicensed}\acmConference[ICMVA 2024]{2024 The 7th International Conference on Machine Vision and Applications}{March 12--14, 2024}{Singapore, Singapore}
\acmBooktitle{2024 The 7th International Conference on Machine Vision and Applications (ICMVA 2024), March 12--14, 2024, Singapore, Singapore}
\acmDOI{10.1145/3653946.3653966}
\acmISBN{979-8-4007-1655-3/24/03}

\renewcommand\footnotetextcopyrightpermission[1]{} 
\AtBeginDocument{%
  }

\begin{document}

\title{Survival Prediction Across Diverse Cancer Types Using Neural Networks\\ 
	}

\author{Xu Yan}
\affiliation{%
  \institution{Trine University}
 \country{USA}
  }
\email{yancontinue@gmail.com}

\author{Weimin Wang}
\affiliation{%
  \institution{Hong Kong University of Science and Technology }
   \country{China}
  }
\email{wangwaynemin@gmail.com}

\author{MingXuan Xiao}
\affiliation{%
  \institution{SouthWest JiaoTong University}
   \country{China}
  }
\email{553556963albert@gmail.com}

\author{Yufeng Li}
\affiliation{%
  \institution{University of Southampton}
   \country{UK}
  }
\email{liyufeng0913@gmail.com}

\author{Min Gao }
\affiliation{%
 \institution{Trine University}
  \country{USA}
  }
\email{mingao4460@gmail.com}

\renewcommand{\shortauthors}{}

\begin{abstract}

Gastric cancer and Colon adenocarcinoma represent widespread and challenging malignancies with high mortality rates and complex treatment landscapes. In response to the critical need for accurate prognosis in cancer patients, the medical community has embraced the 5-year survival rate as a vital metric for estimating patient outcomes. This study introduces a pioneering approach to enhance survival prediction models for gastric and Colon adenocarcinoma patients. Leveraging advanced image analysis techniques, we sliced whole slide images (WSI) of these cancers, extracting comprehensive features to capture nuanced tumor characteristics. Subsequently, we constructed patient-level graphs, encapsulating intricate spatial relationships within tumor tissues. These graphs served as inputs for a sophisticated 4-layer graph convolutional neural network (GCN), designed to exploit the inherent connectivity of the data for comprehensive analysis and prediction. By integrating patients' total survival time and survival status, we computed C-index values for gastric cancer and Colon adenocarcinoma, yielding 0.57 and 0.64, respectively. Significantly surpassing previous convolutional neural network models, these results underscore the efficacy of our approach in accurately predicting patient survival outcomes. This research holds profound implications for both the medical and AI communities, offering insights into cancer biology and progression while advancing personalized treatment strategies. Ultimately, our study represents a significant stride in leveraging AI-driven methodologies to revolutionize cancer prognosis and improve patient outcomes on a global scale.
\end{abstract}



\maketitle
	\section{Introduction}
In the contemporary healthcare landscape, the scourge of cancer continues to loom large, with its pervasive impact felt across the globe. Among the myriad of cancer types, gastric cancer and colon adenocarcinoma stand out due to their widespread prevalence, high mortality rates, and the intricate complexity of their treatment protocols. These malignancies not only pose a significant threat to patient life but also present formidable challenges for medical practitioners striving to combat them effectively. The grim reality is that these cancers are ranked among the leading causes of mortality worldwide, underscoring the urgent need for advanced diagnostic and prognostic tools that can guide clinical decision-making and improve patient outcomes.

The critical importance of accurate cancer prognosis cannot be overstated, as it forms the cornerstone of personalized medicine, enabling clinicians to devise treatment plans that are specifically tailored to the unique characteristics of each patient's cancer. Prognostic assessments play a vital role in determining the course of treatment, ranging from surgical interventions and chemotherapy to targeted therapies and palliative care\cite{10.1242/jcs.116392}\cite{saltz2018spatial}. By accurately predicting the likely progression of the disease, physicians can optimize treatment regimens, mitigate potential side effects, and enhance the quality of life for cancer patients. Moreover, an effective prognosis aids in the allocation of medical resources, ensuring that patients who are most likely to benefit from aggressive treatments receive the attention they need, while also identifying those for whom palliative care would be more appropriate.

Despite the advances in medical science and the development of innovative treatment methodologies, the prediction of cancer outcomes remains a daunting challenge. Traditional prognostic models often rely on a limited set of clinical and histopathological parameters, which, while useful, do not fully capture the complex biological and molecular interactions that drive cancer progression. Furthermore, the heterogeneity of tumors, even within the same cancer type, adds another layer of complexity to the prognosis, making it difficult to achieve a high degree of accuracy in predicting patient outcomes.

Recognizing these challenges, the present study seeks to address the limitations of conventional prognostic models by introducing a novel approach that leverages the power of artificial intelligence (AI) and advanced imaging analysis\cite{10440308}\cite{liu2023unveiling}. By extracting detailed features from whole slide images of gastric and colon adenocarcinomas, and constructing patient-level graphs that encapsulate the intricate spatial relationships within tumor tissues, we aim to provide a more nuanced and comprehensive understanding of tumor characteristics. This, in turn, enables the development of a sophisticated graph convolutional neural network model that harnesses the connectivity of the data to offer precise and reliable predictions of patient survival outcomes.

As we delve into this research, it is our hope that the methodologies and findings presented herein will not only contribute to the enhancement of cancer prognosis but also serve as a beacon for future studies seeking to harness AI and machine learning technologies in the fight against cancer. By pushing the boundaries of what is currently possible in cancer prognostics, we endeavor to pave the way for more personalized, effective, and compassionate care for patients facing the daunting challenge of gastric and colon adenocarcinomas.

	\section{RELATED WORK}
The advent of deep learning has sparked significant advancements in cancer prognosis, particularly in leveraging sophisticated computational models to analyze complex biomedical data. Notably, the application of deep learning techniques in survival analysis of Whole Slide Images (WSI) has garnered considerable attention. Existing methodologies encompass diverse approaches, such as incorporating COX proportional hazard functions in neural networks for survival prediction\cite{Che_Liu_Li_Huang_Hu_2023}\cite{lu2018nuclear}, and leveraging clustering techniques like K-Means at the WSI level to inform convolutional neural network predictions \cite{8100208}\cite{wu2023fineehr}.

However, amidst these advancements, the potential of graph convolutional neural networks (GCNs)\cite{hu2022design} \cite{liu2023ising} stands out as a transformative approach to cancer prognosis. This paper underscores the significance of employing GCNs as the primary model for predicting the survival outcomes of gastric cancer and Colon adenocarcinoma patients. By abstracting WSI slices into graph structures, GCNs offer a novel paradigm for analyzing cancer pathology data. Unlike traditional deep learning methods, GCNs not only capture the features of individual slices within the WSI but also integrate information from adjacent slices, facilitating enhanced perception of the tumor microenvironment and its surrounding context.

The adoption of GCNs represents a paradigm shift in cancer prognosis, offering unparalleled capabilities to extract intricate spatial dependencies and interactions from WSI data. This approach not only enhances the accuracy of survival prediction models but also sheds light on previously unexplored aspects of cancer biology and progression\cite{ma2023implementation}\cite{guo2021multi}. Moreover, in the United States, where cancer prevalence and mortality rates remain significant, the integration of advanced AI techniques like GCNs holds immense promise for improving patient outcomes, optimizing healthcare resource allocation, and advancing precision oncology initiatives\cite{ye2023medlens}\cite{zhang2021distractor}. Thus, this research direction underscores the critical importance of embracing innovative AI-driven methodologies to address the multifaceted challenges posed by cancer in the modern healthcare landscape.

 \section{Methodology}
  \subsection {Graph Convolutional Neural Network}
Before the introduction of Graph Convolutional Neural Networks (GCN), deep learning was predominantly based on Convolutional Neural Networks (CNN). The core of CNN lies in the convolutional kernels performing operations similar to dot products by shifting on input images for feature extraction\cite{yang2017visualization}\cite{dai2023diabetic}. In contrast, GCN integrates features between nodes by connections on the graph's nodes and edges. GCN to a certain extent achieves feature fusion among various node features.

The input of the GCN model is graph data\cite{xu2013union}\cite{lu2023transflow}. Assuming each image has N nodes and M edges, each node has D-dimensional features, then each node will have a D × D-dimensional feature matrix X, and the adjacency matrix A and feature matrix X of N nodes form the input of GCN. The GCN network used in this paper consists of 4 layers, with each layer representing a round of feature learning process.

\subsection{C-index}
C-index, also known as Concordance Index \cite{harrell1996multivariable}, was originally proposed by Professor Frank E Harrell Jr. of Vanderbilt University in 1996 and is commonly used to evaluate the predictive results of survival models. The calculation method of C-index is as follows: pairwise compare among n patients, with a total of C(n, 2) pairs, and then divide the number of pairs where the predicted survival probability is consistent with the actual survival status by C(n, 2). The proportion obtained is the concordance index, which essentially calculates the probability of consistency between the predicted results and the actual status. According to the calculation method above, the value of the concordance index should be between 0.5 and 1.
 
    \section{Experimentation}

    \subsection{Data source} 
The dataset used in this experiment consists of Whole Slide Image (WSI) data for gastric cancer (STAD) and Colon adenocarcinoma (COAD). All the data mentioned above are obtained from the website https://portal.gdc.cancer.gov/.

WSI, which stands for Whole Slide Image, refers to images that cover the entire specimen at high resolution, typically at the level of millions of pixels. The data used in this experiment are WSI images of gastric cancer and Colon adenocarcinoma. Since the research problem addressed in this paper is the 1-year survival rate of gastric cancer and Colon adenocarcinoma, WSI images of cancer patients and survival information are utilized. Each cancer patient may have multiple WSI images, and each cancer patient has their own survival information, including: Overall Survival time (OS.time), which refers to the time from the diagnosis of cancer patients to the last follow-up; Overall Survival status (OS), where OS is either 1 or 0, with 1 indicating the patient's death and 0 indicating the patient's survival; Survival time refers to the time from diagnosis to the last follow-up, measured in months. The final label for each patient is determined based on OS.time and OS, as shown in Figure 1.
\begin{figure}
    \centering
    \includegraphics[width=0.5\linewidth]{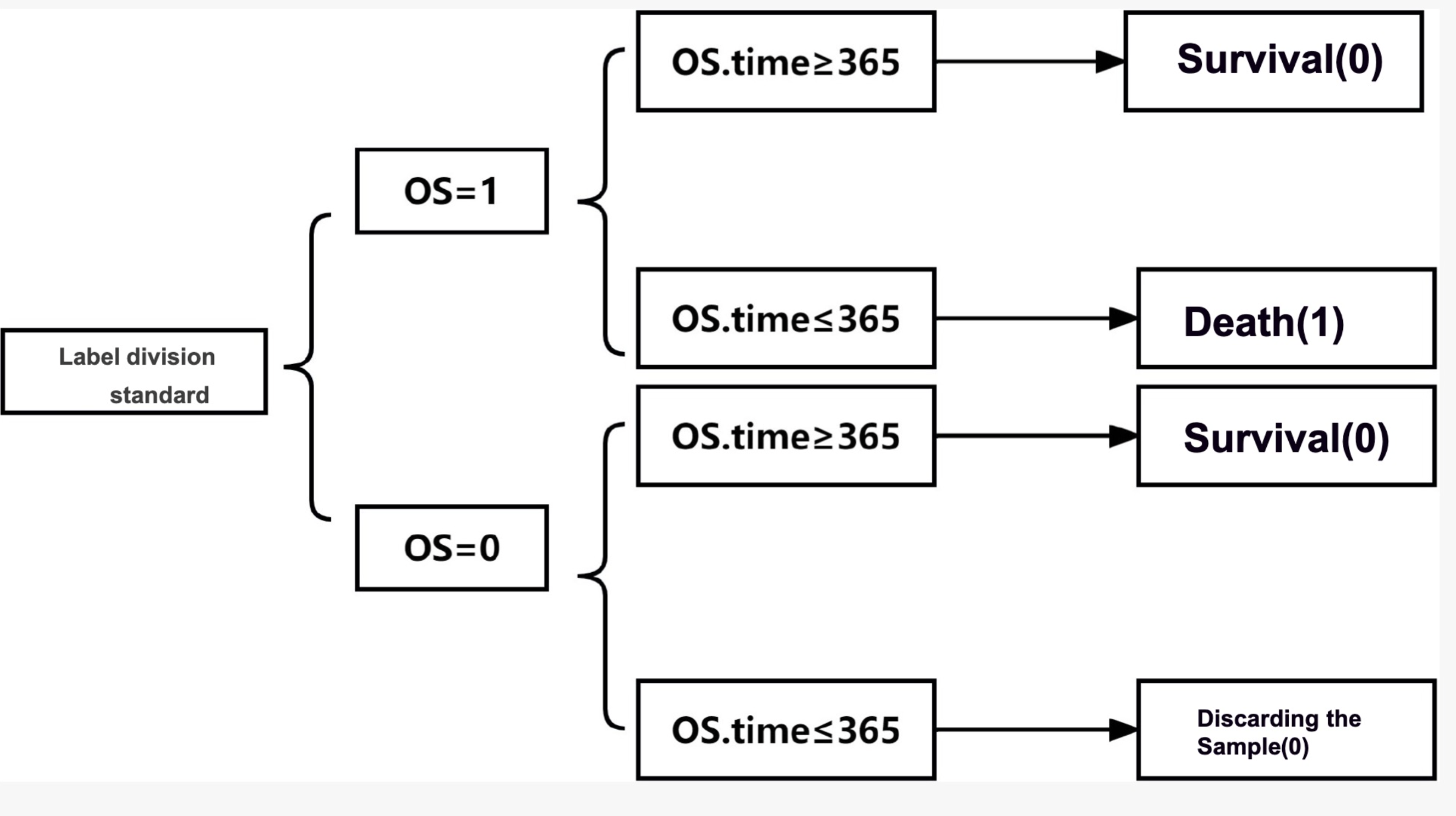}
    \caption{Criteria for dividing patient survival}
    \label{fig:enter-label}
\end{figure}

\subsection{Data Preprocessing} 
For the WSI images of gastric cancer and Colon adenocarcinoma, as shown in Figure 2, we conducted the following preprocessing steps:

\begin{figure}
    \centering
    \includegraphics[width=0.5\linewidth]{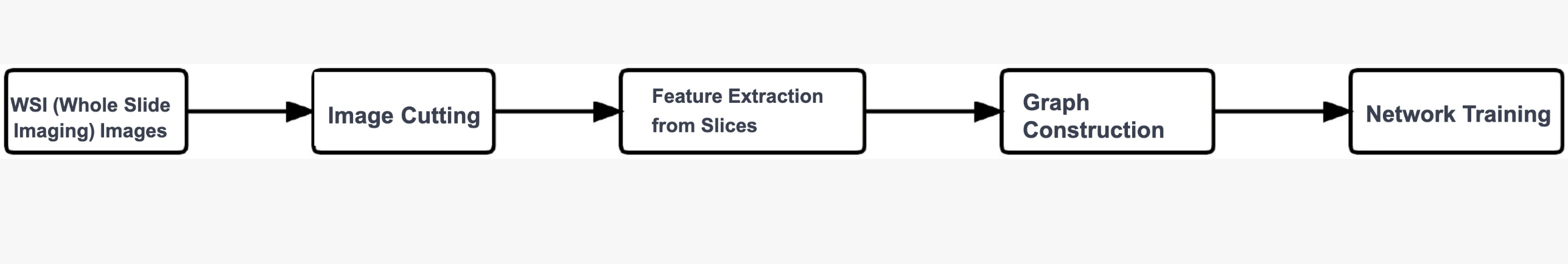}
    \caption{Preprocessing flowchart}
    \label{fig:enter-label}
\end{figure}

WSI (Whole Slide Images), also known as whole slide digital slices, are typically obtained by scanning pathological images of tumors. The pixels of WSI images are extremely large, usually in the range of tens of megabytes to several gigabytes. A single WSI image contains a wealth of pathological information. As shown in Figure 3, this is a WSI image of Colon adenocarcinoma.

\begin{figure}
    \centering
    \includegraphics[width=0.5\linewidth]{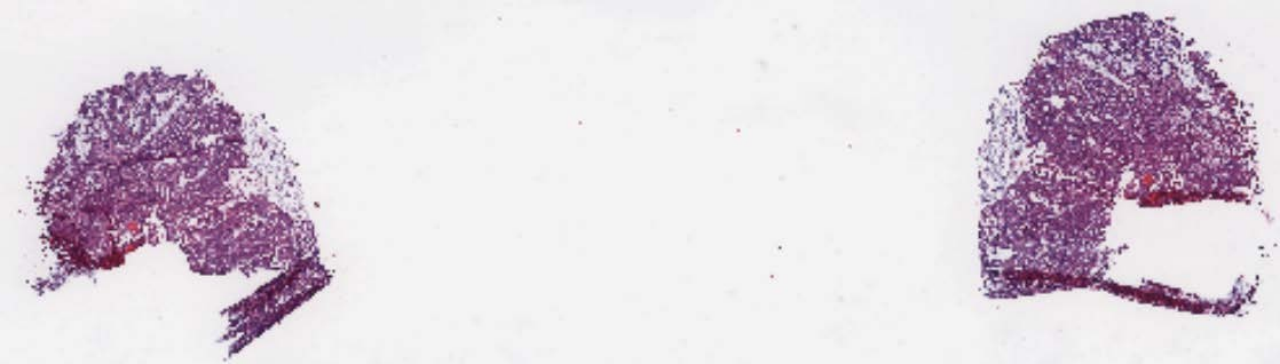}
    \caption{WSI image of Colon adenocarcinoma}
    \label{fig:enter-label}
\end{figure}

After segmenting the WSI, we extracted 1024-dimensional features from the sliced WSI images using the ResNet50 network, which was pre-trained on the ImageNet dataset.

We construct the Graph of WSI at the patient level. For several slices of a WSI, these slices are defined as points on the 2D coordinates of the WSI, forming the nodes of the WSI graph. The 1024-dimensional features extracted by the ResNet network are used as the features for each node. Each slice is connected to its surrounding 8 points on the 2D coordinate plane, forming the edges of the WSI graph.

\subsection{Construction of Graphs}
As shown in the diagram below, taking the red node as an example, it is connected to the surrounding 8 nodes, forming the edges of the entire WSI image. A patient may have multiple WSI images, and the construction shown in the diagram below is applied to each of the patient's WSI images. Subsequently, the graphs formed by these WSI images are combined into a single graph, representing the Graph of the patient. Figure 4 illustrates an example of the construction of nodes and edges for WSI images.

\begin{figure}
    \centering
    \includegraphics[width=0.5\linewidth]{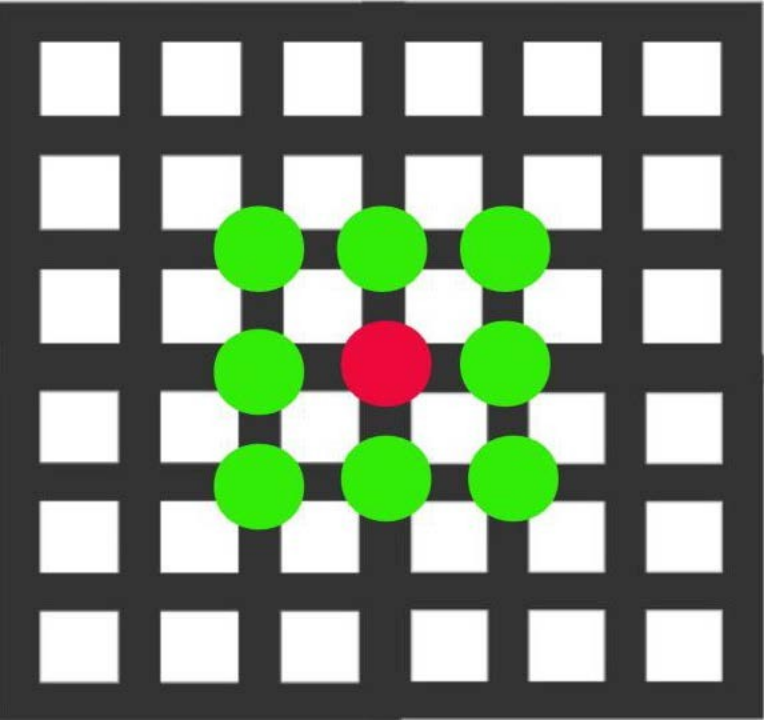}
    \caption{Illustration of the WSI diagram}
    \label{fig:enter-label}
\end{figure}

\subsection{Network Training}
We utilized the Patch-GCN \cite{li2018graph} network for training. Due to the high density of edges resulting from connecting each node on every WSI image to its surrounding 8 nodes, increasing the number of layers in the network imposes a heavy burden on training\cite{10271084}. Additionally, due to the propagation mechanism of GCN, as the number of layers in the network increases, the aggregation of moral information for each node also increases redundantly, resulting in wasted training resources\cite{Hu_Li_Huang_Liu_Che_2023}\cite{liang2022global}. Considering these factors, we set the number of layers in the graph convolutional neural network to 4 layers. Cox function is used for regression in the final layer of the network to perform survival analysis.

\subsection{Experimental Results}    
      We constructed graphs and trained graph convolutional neural networks using data from Stomach adenocarcinoma(STAD) cancer and Colon adenocarcinoma(COAD). The data included WSI image data and 5-year survival data for both types of cancer patients. The patient-level data were divided into training and testing sets with a ratio of 4:1 for 5-fold cross-validation. The C-index values of the five trained models for each cancer were averaged to obtain the final C-index value. Additionally, ROC curves for both types of cancer were plotted.       
        \begin{table}[h]
        \centering
        \begin{tabular}{|p{3cm}|p{5.5cm}|}
        \hline
        \textbf{Cancer types and Neural Networks Models} & \textbf{C-index } \\
        \hline
        STAD with GCN  & 0.64\\
        \hline
        STAD with CNN  & 0.62 \\
        \hline
        COAD with GCN & 0.57 \\
        \hline
        COAD with CNN  & 0.53 \\
        
        \hline

        \end{tabular}
        \caption{Experimental Results}
        \end{table}

Furthermore, to further illustrate the performance of the graph convolutional neural network (GCN) used in this paper, we also trained a convolutional neural network (CNN) on the raw data (WSI), using the graph CNN network proposed by Ruoyu Li et al. \cite{10.1007/978-3-030-87237-3_33}. As shown in Table 1, the final C-index values obtained using the graph convolutional neural network for predicting gastric cancer and Colon adenocarcinoma were 0.64 and 0.57, respectively. The experimental results demonstrate that the graph convolutional neural network (GCN) used in this paper indeed outperforms the convolutional neural network in predicting the survival probability of gastric cancer and Colon adenocarcinoma.

In addition, we plotted the AUC curves for the predictions of the two cancers, as shown in Figure 5 and Figure 6.
\begin{figure}
    \centering
    \includegraphics[width=0.5\linewidth]{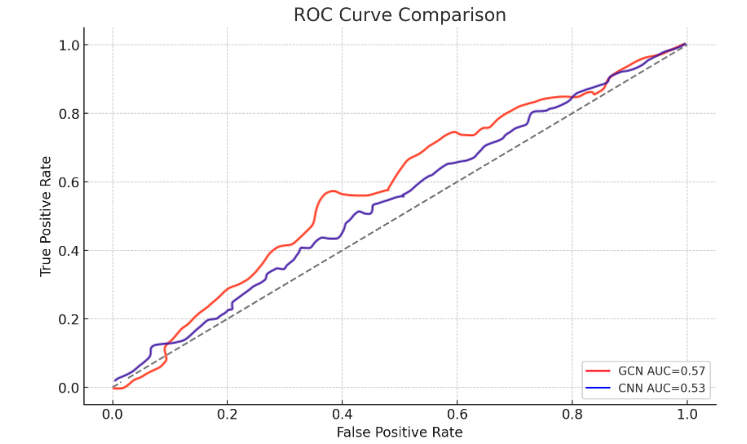}
    \caption{AUC curve for predicting survival probability of Colon adenocarcinoma (COAD)}
    \label{fig:enter-label}
\end{figure}

\begin{figure}
    \centering
    \includegraphics[width=0.5\linewidth]{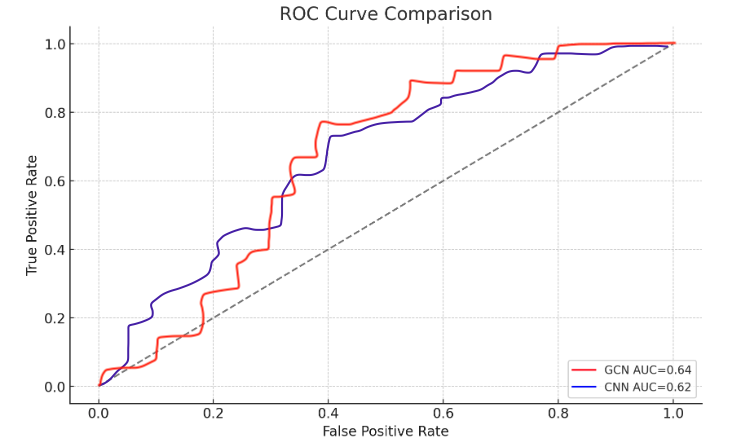}
    \caption{AUC curve for predicting survival probability of gastric cancer (STAD)}
    \label{fig:enter-label}
\end{figure}

        \section{Conclusion}
 
 In this study, we have successfully demonstrated the effectiveness of our approach in predicting the five-year survival information of gastric cancer (STAD) and Colon adenocarcinoma (COAD) patients. By performing preprocessing operations, including segmentation and feature extraction, on Whole Slide Images (WSI), we have enhanced the quality of input data for subsequent analysis. The construction of graphs for each WSI allowed us to capture the complex relationships between different regions within the tissue samples, providing a more comprehensive representation of the pathological characteristics.

Our utilization of graph convolutional neural networks (GCNs) for survival analysis yielded promising results, with C-index values of 0.64 for gastric cancer and 0.57 for Colon adenocarcinoma. These values demonstrate significant improvement compared to traditional convolutional neural networks (CNNs), indicating the superiority of our approach in capturing the intricate spatial dependencies and interactions present in WSI data.

In previous survival analyses of WSI images, most studies used
Convolutional Neural Networks (CNN) for training and obtaining
results. Common methods include Multiple Instance Learning (MIL)
and weakly supervised methods\cite{10.1145/3627377.3627382}. Although these methods can solve many classification and regression tasks on WSI\cite{xu2023characterizing}, they do not achieve "global" learning of WSI. In other words, previous methods did not
integrate the features between different patches in WSI for learning.
However, the Graph Convolutional Neural Network (GCN) has
improved this aspect by utilizing the graph structure at the WSI
level, treating each slice as a node in the graph, completing feature
learning between nodes through connections between nodes, and
thus achieving global learning, to some extent, it can also be seen
as "multi-scale" learning.

The contributions of our research extend beyond the field of medical oncology to the broader realm of artificial intelligence (AI) and machine learning\cite{xu2019can}\cite{10135464}. By leveraging advanced techniques such as GCNs, we have showcased the potential of graph-based models in biomedical image analysis. Our methodology not only enhances the accuracy of survival prediction models but also opens up new avenues for utilizing graph-based architectures in various medical imaging tasks.

Furthermore, our findings hold significant implications for personalized medicine and clinical decision-making. Accurate prediction of cancer survival probabilities enables clinicians to tailor treatment strategies to individual patients, leading to improved patient outcomes and better allocation of healthcare resources. Our research represents a crucial step towards harnessing the power of AI to revolutionize cancer care and underscores the importance of interdisciplinary collaborations between medicine and AI research communities.

       \bibliographystyle{ACM-Reference-Format}

        \bibliography{references}

\end{document}